\documentclass[sigconf]{acmart}

\usepackage{colortbl}
\usepackage{soul}
\usepackage{environ}
\usepackage{mdframed}
\usepackage{hyperref}
\usepackage{multirow}
\usepackage{enumerate}
\usepackage{tcolorbox}

\AtBeginDocument{%
  }

\setcopyright{acmcopyright}
\copyrightyear{2025}
\acmYear{2025}
\acmDOI{XXXXXXX.XXXXXXX}

\acmConference[EASE 2025]{The 29th International Conference on Evaluation and Assessment in Software Engineering}{17–20 June, 2025}{Istanbul, Türkiye}
\acmBooktitle{Proceedings of the 28th International Conference on Evaluation and Assessment in Software Engineering (EASE 2025), June 18--21, 2025, Salerno, Italy}
\acmISBN{978-1-4503-XXXX-X/18/06}

\begin{document}

\newcounter{FindingsCounter}
\NewEnviron{finding}[1][]{%
	\vspace{0.2cm}
	\refstepcounter{FindingsCounter}
	
    \begin{mdframed}[backgroundcolor=white]
        \textbf{Finding \arabic{FindingsCounter}}: \BODY
    \end{mdframed}
    \vspace{0.2cm}
}

\NewEnviron{researchquestion}[1][]{%
	\vspace{0.2cm}
    \begin{mdframed}[backgroundcolor=white]
        \BODY
    \end{mdframed}
    \vspace{0.2cm}
}

\newcounter{StepsCounter}
\NewEnviron{step}[1][]{%
	\refstepcounter{StepsCounter}
    \textbf{Step \arabic{StepsCounter}: }\BODY
}

\newcommand{\datasetURL}{\url{example.com}}
\newcommand{\datasetName}{PRemo}
\newcommand{\datasetMsgNum}{1,791}
\newcommand{\datasetValidatorsNum}{19}
\newcommand{\datasetSysNum}{36}

\newcommand{\citetodo}{\hl{[REF]}}

\newcommand{\numToolsEvaluated}{5}
\newcommand{\la}[1]{{\textcolor{red}{~[~\textbf{LA}: \textit{#1} ]}}}
\newcommand{\wk}[1]{{\textcolor{blue}{~[~\textbf{WK}: \textit{#1} ]}}}
\newcommand{\nc}[1]{{\textcolor{brown}{~[~\textbf{TO BE REMOVED}: \textit{#1} ]}}}
\newcommand{\cm}[1]{{\textcolor{purple}{~[\textit{#1} ]}}}

\title{Relating Complexity, Explicitness, Effectiveness of Refactorings and Non-Functional Requirements: A Replication Study}



\author{Vinícius Soares}
\affiliation{%
  \institution{Pontifical Catholic University of Rio de Janeiro}
  \city{Rio de Janeiro}
  \country{Brazil}
}

\author{Lawrence Arkoh}
\orcid{0009-0005-5904-9313}
\affiliation{%
  \institution{North Carolina State University}
  \city{Raleigh}
  \country{U.S.A}
}

\author{Paulo Roberto Farah}
\orcid{0000-0001-8023-7611}
\affiliation{%
  \institution{Santa Catarina State University}
  \city{ Ibirama}
  \country{Brazil}
}

\author{Anderson Uchôa}
\orcid{0000-0002-6847-5569}
\affiliation{%
  \institution{Federal University of Ceará}
  \city{Itapajé}
  \country{Brazil}
}

\author{Alessandro Garcia}
\orcid{0000-0001-5788-5215}
\affiliation{%
  \institution{Pontifical Catholic University of Rio de Janeiro}
  \city{Rio de Janeiro}
  \country{Brazil}
}

\author{Wesley K. G. Assunção}
\orcid{0000-0002-7557-9091}
\affiliation{%
  \institution{North Carolina State University}
  \city{Raleigh}
  \country{U.S.A}
}

\renewcommand{\shortauthors}{Soares et al.}

\begin{abstract}
Refactoring is a practice widely adopted during software maintenance and evolution. Due to its importance, there is extensive work on the effectiveness of refactoring in achieving code quality. However, developer's intentions are usually overlooked. A more recent area of study involves the concept of self-affirmed refactoring (SAR), where developers explicitly state their intent to refactor. While studies on SAR have made valuable contributions, they provide little insights into refactoring complexity and effectiveness, as well as the refactorings' relations to specific non-functional requirements. A study by Soares et al. addressed such aspects, but it relied on a quite small sample of studied subject systems and refactoring instances. Following the empirical method of replication, we expanded the scope of Soares et al.'s study by doubling the number of projects analyzed and a significantly larger set of validated refactorings (8,408). Our findings only partially align with the original study. We observed that when developers explicitly state their refactoring intent, the resulting changes typically involve a combination of different refactoring types, making them more complex. Additionally, we confirmed that such complex refactorings positively impact code's internal quality attributes. While refactorings aimed at non-functional requirements tend to improve code quality, our findings only partially align with the original study and contradict it in several ways. Notably, SARs often result in fewer negative impacts on internal quality attributes despite their frequent complexity. These insights suggest the importance of simplifying refactorings where possible and explicitly stating their goals, as clear intent helps shape more effective and targeted refactoring strategies.
\end{abstract}

\ccsdesc[500]{Software and its engineering~Extra-functional properties}
\ccsdesc[500]{Software and its engineering~Maintaining software}
\ccsdesc[500]{General and reference~Empirical studies}

\keywords{Software maintenance, Quality attributes, Developer intentions, Empirical study} 


\maketitle

\section{Introduction} \label{sec:introduction}
Refactoring is a key aspect of software maintenance, involving the application of one or more code transformations to enhance quality, without changing the program's behavior~\cite{23}. The process of refactoring is considered effective when it successfully improves internal quality attributes~\cite{11, 23, 27, 40}, such as enhancing cohesion or reducing coupling, complexity and size, which are standard attributes for revealing problematic code~\cite{2}. Furthermore, empirical studies have shown that these internal quality attributes also have impact on various non-functional requirements (or concerns) such as maintainability, security, performance and robustness ~\cite{abid2020does,10536085, traini2021software,accEvavaluate24}.

Despite the wide adoption of refactoring in practice~\cite{34, 101145, barwell2024towards}, recent studies have shown that code refactoring is not always effective in terms of improving internal quality attributes~\cite{8,9,50,OUNI2023111817}. These studies primarily focus on analyzing:
(i) what are the refactoring's main characteristics, i.e., which types of refactoring is being used (e.g., Extract, Move or Pull up Method), and (ii) whether these refactorings are effective (or not) to improve internal quality attributes or to remove the so-called code smells.
However, there are other key refactoring characteristics that still need investigation to determine to which extent they relate to refactoring effectiveness. These characteristics include \textit{refactoring complexity} as well as  \textit{developer explicitness} on refactoring activities and which \textit{non-functional requirements} are targeted by such refactorings. 

Being explicit means that the developer has a clear intention about the act of refactoring and documenting such an intention in commit messages. This explicitness has been widely termed as Self-Affirmed Refactoring (SAR)~\cite{3}. 
Independently of the intention, refactoring is a complex task.  
Studies have shown that developers apply complex refactoring in the form of multiple inter-related refactoring types on single or multiple code elements, which is known as \textit{batch} or \textit{composite refactoring}~\cite{8, 50, BIBIANO2023107134}. These refactorings can range from a simple, repeated usage of one or two different transformation types, to complex processes spanning over five different transformation types~\cite{9}. Some empirical studies have shown the relation between composite refactorings and their effectiveness on the improvement of code quality~\cite{8, 34, 47, BIBIANO2023107134, 10795281, agnihotriImpabatch, rego2018understanding, imran2024towards}. 


When refactoring, developers can also have a variety of different concerns, including the improvement of system's conformity to \textit{Non-Functional Requirements (NFRs)}~\cite{30}. NFRs are the requirements that do not dictate what a software system should do, but rather define software properties or constraints under which the software should operate~\cite{glinz2007non, 101145nfr, chung2012non}. NFRs address important software quality aspects such as performance, maintainability, robustness and security~\cite{chung2012non}. As mentioned earlier, NFRs are closely related to internal quality attributes, which in turn, are directly influenced by refactoring. Thus, to better understand refactorings, these should be explicitly associated with the NFR(s) they aim to impact. 


A study by Soares et al.~\cite{47} (also referred hereafter as \textit{original work}) has advanced the state-of-the-art on investigating the relations among refactorings' effectiveness, their complexity as well as their relations with SARs and NFRs. This study suggests that developers apply more complex refactorings when they explicitly manifested their concerns with NFRs, and that refactoring may have recurring effects when paired with a specific NFR concern~\cite{47}. For instance, the study indicates that explicit concerns with security and performance have negative effects on internal quality attributes. However, despite the promising findings, the original work presents various shortcomings, including (but not limited to) the reduced number of projects, much smaller refactoring dataset, lack of statistical tests, and lack of discussion about the findings' implications.

The study reported in this present paper is a replication of Soares et al.'s study~\cite{47}. 
To this end, we analyzed a set comprised of four projects from the original work as well as four other new projects, thus, doubling the size of the original work's dataset in terms of project count. Alongside this, due to the upgrade to RefactoringMiner 2.0~\cite{53} and the updates to improve the methodology, we have more than tripled the total number of analyzed refactorings of the original work, obtaining a total of 8,408 refactorings. We have obtained information on both these refactoring's effects on internal quality attributes, and their complexity (number of unique transformation types). To identify SAR, we used the same keyword-based classification method as the original work, with the same keyword list, which maintained a relatively high F1-score of 78\% even with the addition of the new projects. Through this analysis, we present the following findings: 

\textbf{Refactoring Complexity:} Similar to the original work, we are able to confirm that the complexity of refactoring has a correlation to its effectiveness in improving internal quality attributes. We found that as refactoring complexity increases, there is an increase in positive effects on internal quality attributes. \textit{Contribution:} This observation reinforces the importance of addressing refactoring complexity in practice, showing that it is a factor that should be considered when designing and analyzing refactoring techniques. With proper support (e.g., refactoring recommenders), the application of complex refactorings may be able to achieve more positive results than the non-positive effects.

\textbf{Refactoring Explicitness: } We also confirm some of the results on the explicitness, as our findings show that refactoring in which developers showed explicit concerns with the refactoring process (SAR~\cite{3}) resulted in frequent and complex refactorings. However, unlike the original work, we observed a decrease in the negative effects of such explicit refactorings on internal quality attributes. That is, negative effects are less common in refactorings in which developers manifest their intentions (i.e., SARs). This shows that developers may perform refactorings more effectively if they are primarily concerned with the refactoring process itself. However, this does not exclude the possibility that the lack of tool support can be problematic for complex refactorings, as was also stated by other studies~\cite{30, 55, 51}. \textit{Contribution:} In addition to having tooling support merely suggesting complex refactorings, our results indicate that developers should also be explicit in their intentions when refactoring. 
While SAR results in complex refactoring, they are inherently not detrimental to internal quality and hence should be encouraged in software maintenance.

\textbf{Non-Functional Concerns: } We were unable to replicate the results of the original work with regard to NFRs relating to refactoring effectiveness. 
In this work, we observed that when developers are concerned with NFRs, especially with maintainability, performance, and security, there is an increase in positive effects and a decrease in negative effects. Our findings also show that refactorings with maintainability and performance have balanced positive improvement across the four code quality attributes (i.e., cohesion,  complexity, coupling and size), while security and robustness have positive improvements in one or two specific attributes.
\textit{Contribution:} Our results suggest that developers concerned with the quality of code, in its many aspects, perform more directed and effective changes. Yet, we reinforce the need for deeper analysis of potential correlations between refactoring applications and NFRs. These further analyses can focus on using large datasets, potentially leading to interesting results, since some NFRs have specific quality attributes as targets.


\section{Related Work} \label{sec:related_work}

This section provides an overview of existing literature related to our study and the original work. We organize the related work into the three following groups: effectiveness, explicitness, and non-functional requirements. 

\textbf{Effectiveness:}~\citet{5} reports an empirical analysis of three Java systems in order to evaluate claims that refactoring improves software quality. The study evaluates possible correlations of internal quality attributes of cohesion, coupling, code complexity, inheritance, and size with external quality attributes such as maintainability and testability. The author concludes that refactorings seldom had a positive impact on these attributes, tending to be mostly neutral. However, a study by~\citet{9263167} suggests that refactoring, when done properly, reduces code coupling.~\citet{Almogahed10244} performed an impact assessment of 10 refactoring types on five projects using the Quality Model for Object-Oriented Design
(QMOOD). These authors observed that refactoring methods generally have positive improvement effects on cohesion, design size, and complexity, even though there are instances where negative and neutral effects of those refactorings were recorded.
\citet{17} investigate how root-canal and floss refactoring~\cite{murphy2009we, noei2023empirical} related to internal quality attributes, including cohesion, coupling, code complexity, and inheritance. They find that over 94\% of applied refactorings negatively impact at least one internal quality attribute. Moreover, most refactorings improve their quality attributes while others keep them unaffected. Another contrasting result stated that about 93\% of root-canal refactoring led to significant improvement in the software's quality attributes like coupling between objects, cohesion of methods, and weighted methods per class~\cite{9734141}.

In the context of complex refactorings,~\citet{9} performed a study on different open-source projects in order to analyze the impact of incomplete composite refactorings in their quality attributes. Incomplete composite refactorings are those that do not fully remove code smells~\cite{10}. They found that most incomplete composites have a neutral effect on internal quality, neither increasing nor decreasing code quality. Also, in the context of composite refactorings,~\citet{21} evaluated how composite refactoring compares to single refactorings in improving cohesion, code complexity, coupling, inheritance, and size of affected elements. Among the analyzed refactorings, 65\% improved attributes associated with their refactoring types, while 35\% kept them unaffected. A study by~\citet{agnihotriImpabatch} reveals that composite refactorings are very effective in improving code quality attributes without degradation or side-effects. 

The aforementioned studies investigate potential correlations between refactoring effectiveness and other factors. However, they did not consider refactoring complexity and the presence of non-functional requirements as potential factors.

\textbf{Explicitness:} The concept of Self-Affirmed Refactoring (SAR) in software development has gained attention in recent years~\cite{alomar2023state}. \citet{42} proposes a phrase-based approach to detecting developer discussions related to refactoring, which they achieved with high accuracy. Likewise, for SAR-related research, \citet{8870177} performed a study to determine which quality metrics are in line with developers' vision when they explicitly mention their refactoring intentions. They found that several metrics associated with cohesion, coupling, complexity, and design size vary positively as developers explicitly refactor to improve them. Nevertheless, another study observed that when developers refactored mainly with the intent of reusability, the number of methods increases, but this does not result in any improvement in the majority of the state-of-the-art metrics~\cite{Alomar2022aa}. Also, the authors of another study found that commit messages for SARs have more significant refactoring activity~\cite{3}. However, the study did not account for the possible relations of such significant refactoring activity with internal quality attributes or the effectiveness of these refactoring in improving these quality attributes.


\textbf{Non-Functional Requirements (NFRs):} The relationship between internal quality attributes and NFRs has been studied by a number of studies. We focus our analysis on studies addressing refactoring and the NFRs analyzed in this work.

Regarding performance,~\citet{45} analyzed refactoring effects on the performance of software product lines. They found that refactorings such as Inline Method and Inline Class can reduce the execution time on method calls, thus improving performance. Also,~\citet{26} showed that removing code delegation and indirection can improve software performance by around 50\%.~\citet{19} reported performance improvements after replacing conditional statements with call methods through polymorphism.
Some studies also analyzed the relation between size (defined in terms of code statements) and performance~\cite{41, 24}. \citet{46} discussed performance antipatterns based on coupling, cohesion, and code complexity of the inheritance hierarchy, concluding that God classes are detrimental to performance.

In the context of robustness, it is common to use exception flow information~\cite{12, 13, oliveira2023don} with the goal of improving code reliability by providing constructs for sectioning code into exception scopes (e.g., try blocks) and exception handlers (e.g. catch blocks). \citet{28} evaluated the robustness of 50 projects by using the internal quality attributes of size and code complexity. Their finding suggests that exception handlers are usually simplistic and that developers often pay little attention to exception scoping and behaviour handling.

We also found studies relating internal quality attributes to security. \citet{18} evaluates how internal quality metrics of coupling, cohesion and complexity can indicate security risk in software. They conclude that size metrics can indicate structures that could be exploited to cause denial-of-service attacks, while coupling can impact on how damage may propagate to other components of the software. Furthermore, their results show that these metrics are not sufficient to indicate specific vulnerability types. Other studies~\cite{44, 33} also support these findings.
\citet{Abid9130035} studied refactoring effects on 30 open-source Java projects, and they observed that refactoring could either have a negative or positive correlation with the security metrics depending on the type of refactoring being applied.

For maintainability, \citet{9825885} found that the application refactorings may have positive, negative, or neutral effects on code readability, which correlates with the maintainability of software projects. However, another study observed that refactoring types like move method, extract method, and extract class have a positive impact on maintainability and understandability~\cite{8261034}. Despite most studies on the effects of refactoring on maintainability using open-source projects, a study by \citet{7732033} made their analysis using five proprietary software projects with which they analyzed maintainability based on internal quality attributes. They found that refactoring methods such as Extract Method, Consolidate Conditional Expression, and Hide Method increase the level of code maintainability while refactoring types like Encapsulate Field and Extract Class reduce code maintainability.

The identification of NFRs in text and developer discussion has also been explored. \citet{31} proposed an automatic classification of user reviews into concerns with four NFR types, namely reliability, usability, portability and performance. They evaluated the combinations of the classification techniques and machine learning algorithms with user reviews collected from two popular mobile apps from different platforms and domains. Results show that a combination of algorithms achieves an F-measure of 71.8\%. \citet{14} applied a semi-supervised learning approach to identify NFRs in textual specifications using a collection of requirements-related documents from 15 software projects, consisting of 370 mentions of NFRs and 255 functional ones.

While these studies analyzed the applications of refactoring in the context
of NFRs, all of them focused on how refactorings could be used in order to
improve attributes specific to each NFR. Our work, on the other hand, focuses
on how developer concerns with specific NFRs can affect the refactoring’s
effectiveness, both in general and looking at each specific internal quality
attribute.

\section{Study Design} \label{sec:methodology}
As this is a replication study, our primary study is the original work by Soares et al., titled ``On the Relation between Complexity, Explicitness, Effectiveness of Refactorings and Non-Functional Concerns''~\cite{47}. 
Similar to the original study, our main goal is to \textit{investigate to what extent refactoring complexity, as well as quality-related developer concerns (explicit refactoring and NFR concerns), relates to its effectiveness, i.e., the improvement of the software's internal quality}. This goal leads us to seek answers to three research questions (RQs) as follows:

    \textbf{RQ1. Is the complexity of refactoring related to their effectiveness?} As mentioned in the original study, there exists a correlation between refactoring \textit{complexity} and its \textit{effectiveness}; we seek to confirm this correlation and understand how the effectiveness varies with respect to the complexity of the refactoring. We adopted the same meaning for complexity and effectiveness as used in the previous work. Thus, \textit{complexity} refers to \textit{the number of different refactoring types that compose the applied refactoring} while \textit{effectiveness} refers to \textit{the impact of the refactoring in improving internal quality attributes by improving their associated metrics}.

    \textbf{RQ2. Does the explicitness of refactorings relate to its complexity and effectiveness?} The original study observed a correlation between refactorings in which developers explicitly stated their intentions and the refactorings' complexity and effectiveness. Developers being explicit while refactoring means \textit{the presence of SAR in either the commit message, issue, or pull request related to the changes with which refactoring was applied}. Thus, we aim to confirm this correlation and understand how it affects the correlated factors.

    \textbf{RQ3. Are non-functional requirements (NFRs) related to the effectiveness of refactorings?} For this RQ, we seek to confirm the findings of the original work which found a correlation between NFRs and refactoring effectiveness. Similarly to the original study, we define NFRs as \textit{the presence of one of four analyzed NFRs in either a commit message, issue, or pull request (or comment) related to the changes where refactoring was applied}.

The RQs outlined above can be addressed based on key factors which are mainly \textit{Internal Quality Attributes} and \textit{Non-Functional Requirements}. Our selection of these attributes and requirements is based on the original study.
For \textit{Internal Quality Attributes}, we analyze cohesion, code complexity, coupling, and size. These attributes are widely used in other studies~\cite{9}, and are connected to the NFRs chosen for this work (e.g., size and complexity correlate to performance)~\cite{shahid2017impact}.
Regarding \textit{Non-Functional Requirements}, we focus on \textit{maintainability, robustness, performance, and security}. We consider maintainability as the primary concern of refactoring since this is the major aim of code refactoring~\cite{pantiuchina}. Robustness can be improved by restructuring modules to integrate patterns that improve error handling (e.g., Chain of Responsibility~\cite{25}). The performance can be improved by eliminating code redundancies (via refactoring) and fixing the sub-optimal distribution of code entities to reduce execution time~\cite{traini2021software}.

\begin{figure}
    \centering
    \includegraphics[width=1\linewidth]{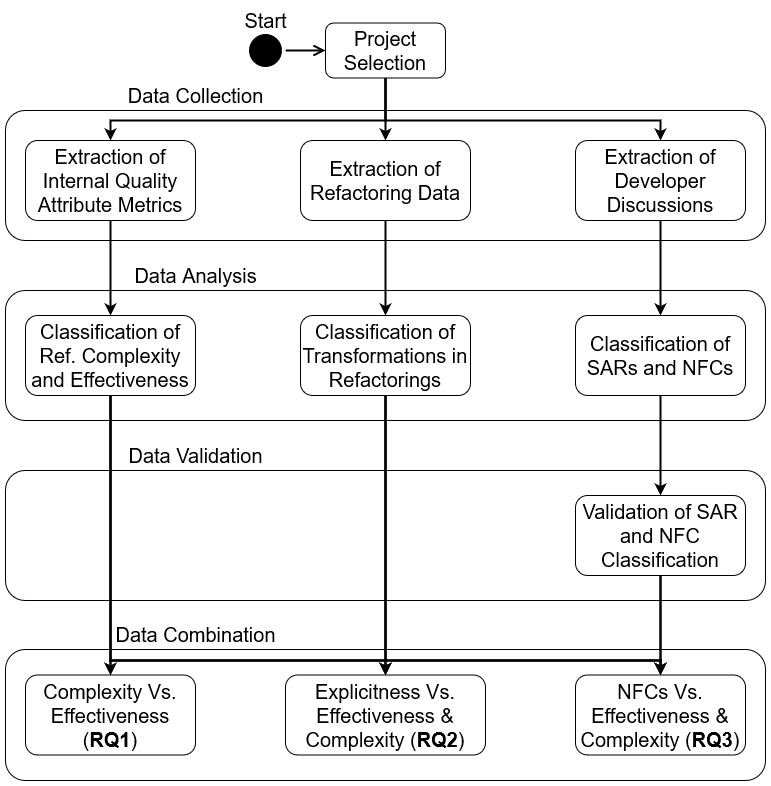}
    \caption{Adopted methodology}
    \label{fig:methodology}
\end{figure}

As mentioned previously, we adopted the methodology of the original work. However, whereas the original work considered only four subject systems, we added four new projects. Additionally, we applied a more strict process to improve the potential statistical accuracy of our work. Figure~\ref{fig:methodology}  shows an overview of our entire methodology. In the following subsections, we describe each step in detail.

\subsection{Project Selection}
The original study selected projects based on three criteria. First, the project must be open source and based on Java programming language. The choice of open source projects improves the replicability of the study, while that of Java relies on the fact that there exists a variety of support from tools for analyzing structural quality. The other reason for choosing Java based projects is that RefactoringMiner tool which we heavily used in our study supports Java projects only\footnote{\url{https://github.com/tsantalis/RefactoringMiner}}. 
Secondly, we considered the number of contributors as well as their code contribution activity. Thus, we selected active projects (i.e., those with frequent and recent code contributions as at the time of data collection) with a considerable number of contributors. This allows us to collect a diverse set of developer discussions and commit messages, which are essential for SAR and NFR analysis.
Lastly, we considered projects which are at least five years old at the time of analysis. The use of such mature projects reveals some level of architectural or code degradation over a period of time, which presents an opportunity for us to analyze developer's refactorings to mitigate such degradations. Also, we select projects that have undergone a large variety of refactoring transformations as defined by~\citet{23}. This allows us to analyze the complexity of refactoring actions with different transformation types.
By applying the above criteria, we selected four new projects which are Fresco, RxJava, Presto and Netty. Also, we kept the existing projects which are Couchbase Java Client, Dubbo, OKHttp and JGit making a total of eight projects for use in our study. The number of refactorings for each project is presented in Table~\ref{tab:projects}.

\begin{table}[!tp]
    \caption{Total number of refactorings per project.}
    \label{tab:projects}
    \centering
    \begin{tabular}{llr}
        \toprule
        \textbf{Source} & \textbf{Project} & \textbf{Refactorings} \\
        \midrule
        \multirow{4}{*}{Orignal Work} & Couchbase & 213 \\
        & Dubbo & 781 \\
        & OKHttp & 689 \\
        & JGit & 1,244 \\
        \midrule
        \multirow{4}{*}{Our study (new projects)} & Netty & 2,225 \\
        & Presto & 1,853 \\
        & RxJava & 912 \\
        & Fresco & 491 \\
        \midrule
        \multicolumn{2}{l}{Total} & 8,408 \\
        \bottomrule
    \end{tabular}
\end{table}

\subsection{Data Collection}
To answer our RQs, we collected data from three different groups: \textit{Structural Quality Data}, \textit{Refactoring Data}, and \textit{Developer Discussions}. We describe the data collection from each group as follows.

\textbf{Structural Quality Data:} This is composed of the internal quality attributes for each code element of the project. We used the Understand\footnote{\url{https://scitools.com/}} static code analysis tool to collect metrics related to internal quality attributes, namely cohesion, code complexity, coupling, and size. By collecting these metrics, we can analyze each pair of subsequent commits in order to determine the changes in quality caused by each commit as either improving or worsening these attributes. Similar to the original study, we only collected a subset of Understand's supported metrics, as listed in our supplementary materials~\cite{soares_2025_14780166}.

\textbf{Refactoring Data:} This data comprises a list of refactoring transformations that have been applied in each commit. We used the RefactoringMiner 2.0 tool~\cite{53} to detect and classify the available refactorings based on the transformations proposed by Fowler et al.~\cite{23}. We chose this tool because it has a high reported precision of 99\%, with a recall of 94\%~\cite{53} making it a reliable tool for refactoring detection and classification.

\textbf{Developer Discussions:} We referred to discussions recorded in the project's repository to gather this data. We used the GitHub API to extract the commit messages, issues, and pull requests from the repositories of the selected projects. Similar to the original study, Couchbase Java Client and JGit did not have Issue and Pull Request data, and thus were only comprised of commit messages. However, SAR detection is still accurate even with this minimal data, as mentioned by Soares et al.'s original work~\cite{47}.

\subsection{Data Analysis}
After we had obtained the data using the above steps, we proceeded with the analysis of the data related to \textit{effectiveness}, \textit{complexity}, \textit{explicitness}, and \textit{non-functional requirements} of refactoring. We describe the process as follows.

\textbf{Refactoring complexity and Effectiveness:} As mentioned previously, we defined refactoring complexity as the number of transformation types used in each refactoring. This has been found to directly relate to the effectiveness of refactorings in the original work~\cite{47}. Also, by using the RefactoringMiner tool, we extracted 40 unique refactoring transformation types at both class and method levels. These include Extract Superclass, Inline Method and Move Attribute. The full list of these transformations is listed on our supplementary material page~\cite{soares_2025_14780166}.

To determine the complexity of these refactorings, we grouped them as single or composite refactoring using the commit-based approach proposed by~\citet{50}. \textit{Composite refactoring} commits refer to those with multiple code transformations within the same commit since these multiple transformations contribute to the same task, while \textit{Single Refactoring} refers to those with single code transformations. The usage of these commits as our unit of measurement also allowed us to determine the relationship between SARs, NFR and the refactoring actions themselves. It is also worth mentioning that SAR and NFR are directly related to single commits.

Regarding refactoring effectiveness, we analyzed the metrics for internal quality attributes such as complexity, cohesion, coupling, and size of code and how they changed between successive refactoring commits. To achieve this, we identified the code elements affected by a given refactoring commit and measured the changes in internal quality attributes between the previous commit and the current refactoring commit. We then classified the change behavior as either positive, negative, or neutral using the following criteria~\cite{47}:
\begin{enumerate}[(i)]
    \item A change is \textit{positive} if at least one of the metrics related to a specific attribute changed positively between the two versions.
    
    \item A change is \textit{negative} if no positive changes occurred, and at least one of the metrics changed negatively.
    
    \item A change is \textit{neutral} if none of the previous two conditions were fulfilled.
\end{enumerate}

By using these criteria for refactoring effectiveness and complexity, we were able to answer RQ1.

\textbf{Presence of Self-Affirmed Refactorings:} We followed the same approach described in the original work to determine the presence of SAR~\cite{47}. We created an automatic, keyword-based classifier that matches a set of 11 keywords and eight key phrases with the developer discussions in order to determine if refactoring intent was discussed or not. The original set of keywords was adapted from Ratzinger's work~\cite{42}. However, this set of words was changed in the original work to improve the classifier accuracy for the projects being studied. 
By combining this data with the previously collected complexity and effectiveness data, we are able to answer RQ2. 

\textbf{Non-Functional Concerns: } In the original work, the authors mentioned that existing NFR detectors and classifiers have very low accuracy~\cite{47}. 
Hence, we performed this action manually. In addition to the existing data obtained in the original work, we manually validated 817 commits, which allowed us to answer RQ3. This validation is described in the next section.

\subsection{Validation}
To determine the trustability of the automatic SAR classification and forming a manually classified data set of NFRs, we performed a manual validation with a subset of commits from the four new projects we included in the replication. The reason we focused on performing a validation on only the four new projects is due to the original work already released its own validated data set, thus allowing us to combine both sets to perform further verifications. The following subsections describe the validation process.

\textbf{Validation Process:} 
For each of our four new projects (see Table~\ref{tab:projects}), we selected a set of 56 commits, making a complete set of 224 different commits. We further grouped these commits into four different groups based on the results from the automatic SAR detection, and RefactoringMiner’s refactorings as follows:

\begin{enumerate}[(i)]
    \item \textbf{+SAR +Ref}: commits classified as SAR, and classified as refactorings by RefactoringMiner.
    
    \item \textbf{+SAR -Ref}: commits classified as SAR, but not classified as refactorings by RefactoringMiner.
    
    \item \textbf{-SAR +Ref}: commits not classified as SAR, but classified as refactorings by RefactoringMiner.

    \item \textbf{-SAR -Ref}:  commits not classified as SAR, and also not classified as refactorings by RefactoringMiner.
\end{enumerate}

Again, we split these refactorings across the four projects to mitigate potentially biased results from focusing too much on a single project.
We performed this validation with seven participants who are knowledgeable in both refactoring and NFRs. We also provided them with documented clear definitions of refactorings, self-affirmed refactorings and NFRs to ensure consistency.
They performed the validation in two steps. First, each participant validated a set of 32 commits, which consists of eight commits from each of the four projects. 
After completing this, we reshuffled the dataset to ensure that no participant had a repeating commit, and then a second round of validations was performed.
After the completion of these two rounds of validations, two authors from the work determined the final category of conflicting classification identified in the previous step based on their personal experience and the confidence level of the answers by the participants.
The participants were asked to identify the
following: 

\begin{enumerate}[(i)]
\item if the commit contained a SAR in any of its discussions.
\item which sentence, and in which location (commit message, issue, etc.) was the SAR located.
\item which keywords in the sentence could be used to detect the SAR.
\item if any NFR was present, and which sentences contained potential NFRs.
\item if a maintainability-related NFR was present.
\item if a robustness-related NFR was present.
\item if a performance-related NFR was present.
\item if a security-related NFR was present.
\end{enumerate}

Finally, for (i), (v), (vi), (vii) and (viii), the participants were asked to determine on a scale of 1 to 5, which confidence level they had in their classification. The full results of this validation are available in the supplementary material~\cite{soares_2025_14780166}.

\textbf{Validation of Self-Affirmed Refactorings:} 
For this validation, we focused on the precision and recall of the SAR detector. We recorded a precision of 69.7\% while the recall was 80\% and the F1-score was 74.5\%. We also attempted to use the Azure-based machine-learning classifier proposed by AlOmar et al. ~\cite{4} to classify our validated dataset for the four new projects. We observed a precision of 52.3\%, a recall of 80.3\% and an F1-score of 63.3\% with this classifier. Thus, we could see that without changing the keyword set, the keyword-based classifier achieves a higher accuracy and hence, we decided to use the same keyword classifier as used in the original study.
By combining our newly validated dataset with that of the original study and using the keyword-based classifier, we recorded a precision of 74.9\%, a recall of 81.4\% and an F1-score of 78\%.

\textbf{Validation of Non-Functional Concerns:} 
We abandoned automatic classification and manually validated the commits for non-functional concerns due to the low accuracy recorded in the original study. Thus, by combining this new classified set with the original validated set released by Soares et al.~\cite{47}, we were able to have the set of 817 classified commits. From this group, 449 commits mentioned NFRs in their discussions, while the remaining 368 commits did not. Considering each individual NFR, we had 197 refactorings related to Maintainability, 126 related to Robustness, 86 related to Performance, and 40 related to Security, including refactorings that can be classified as multiple types.

\section{Results and Discussion} \label{sec:results}
In this section, we present the results of our analysis of the data, which helps us answer the RQs introduced in Section~\ref{sec:methodology}. First, we answer RQ1 by performing an analysis of the potential relationship between refactoring complexity and effectiveness. Then, we perform an analysis of the results of RQ1 with regard to the presence of SARs to answer RQ2. Finally, we then correlate the findings of RQ1 with the presence of NFRs to answer RQ3. While we describe the results in this paper, the raw data and results for each of the described analyses are available on the supplementary material~\cite{soares_2025_14780166}.

\subsection{RQ1: Refactoring Complexity vs. Effectiveness}
\begin{figure}[ht!]
    \centering
    \includegraphics[width=1\linewidth]{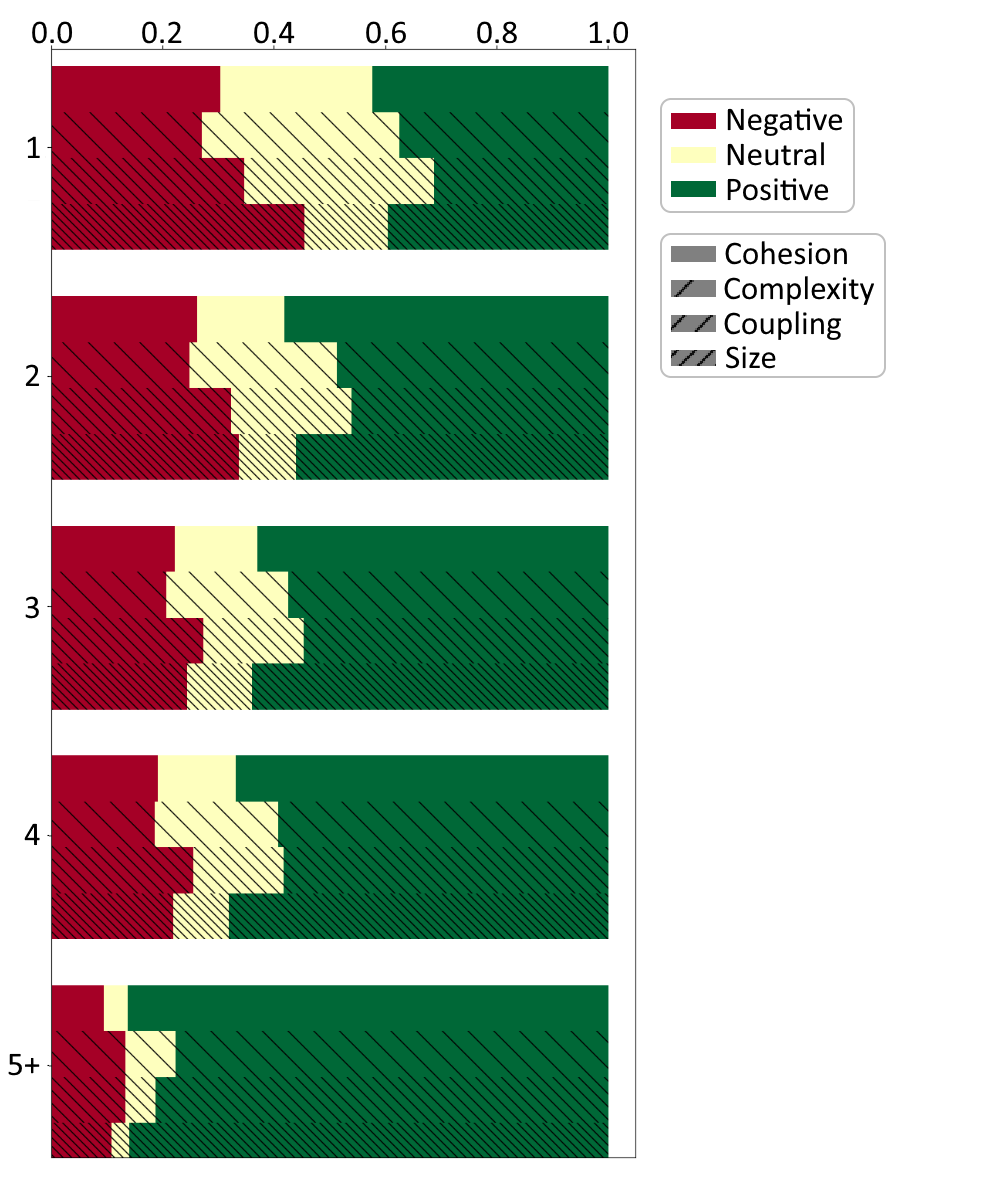}
    \caption{Distribution (decimal percentage) of effects based on the refactoring complexity. Each 0.1 on the horizontal scale represents 10\% in change frequency of the corresponding effectiveness.}
    \label{fig:complexity_vs_effectiveness_and_internal_quality_attributes}
\end{figure}
We grouped refactoring commits into five categories based on the number of different transformation types employed, as in the original work. That is, categories 1 to 4 contained refactorings composed of 1 to 4 transformation types, respectively, whereas category 5+ are those refactorings with five or more transformation types.

Figure~\ref{fig:complexity_vs_effectiveness_and_internal_quality_attributes} presents the distribution of the effectiveness of the various refactoring complexity categories based on cohesion, code complexity, coupling, and code size. The various categories of refactoring complexities are shown on the vertical axis. From this figure, we can see a general reduction in the negative and neutral effects of refactoring as the complexity of the refactoring increases from 1 to 5+. This is also reflected as an increase in positive effects from complexity 1 to 5+. Considering all projects, we recorded that the average percentage of neutral effects reduced from 28\% (at complexity 1) to 5.5\% (at complexity 5+). This represents a very significant reduction as the neutrality of 5+ refactorings is quite close to 0\%. Similarly, the average value for negative effects reduced from 34\% at Category 1 to 11\% at Category 5+. 
From Figure \ref{fig:complexity_vs_effectiveness_and_internal_quality_attributes}, we can see a sharp increase in the positive effects of refactoring as the complexity increases. This means that high-complexity refactorings, i.e., those that combine a high variety of different transformation
types into a single, large refactoring, can have significantly better effects on the code. This observation differs from the original work~\cite{47}, since Soares et al. observed a balanced increase in both positive and negative effects of refactoring as the complexity increases. However, our results give a very clear indication of the increasing positive effects of code quality as refactoring complexity increases. 

Our finding also confirms the results from studies conducted by~\citet{agnihotriImpabatch}  as well as~\citet{21} that composite refactorings (refactorings comprising several transformation types) led to improvement in code quality metrics even though we recorded some degree of degradation. Also, our results confirm that of an existing study ~\cite{OUNI2023111817} in which they observed 16 co-occurent refactorings had a greater positive impact. In this same study, the authors observed that single refactorings have 23.1\% and 2.2\%  positive and negative impact respectively on quality metrics. Our results, however, show relatively greater values for the positive and negative impacts. For complex refactorings, the same authors recorded 42.4\% and 3.3\% positive and negative impacts, respectively. Thus, our results confirm their observation that as the complexity of refactoring increases, the potential for positive effects of these refactorings increases as well. With respect to the negative effects, our observations confirm that of existing studies that refactorings, irrespective of their complexities, do introduce some level of negative effects or code smells~\cite{10795281}.

By using the Kruskal-Wallis test, we found that the statistical
significance for the results found in (code) complexity and coupling are quite
close to significant (p-values < 0.05), suggesting meaningful trends that differentiate these factors across groups
However, the results for cohesion and, especially, size did not achieve the same statistical significance (p-values of 0.10 and 0.19, respectively). Their p-values indicate noteworthy patterns that may provide valuable insights with further exploration. These findings highlight the relevance of complexity and coupling while also suggesting that cohesion and size could contribute to a more comprehensive understanding of such relations.

\begin{tcolorbox}
\textit{\textbf{RQ1 Answer :}
As refactoring complexity increases, the frequency of positive effects on the code also increases with it. This means the possibility of refactorings causing non-positive (harmful) effects is reduced. As aforementioned, the results of this RQ are quite interesting. It only partly confirms the previous study’s findings about refactoring complexity (the number of different transformation types) being related to refactoring effectiveness. We found here that more complex refactorings tend to have a clear positive effect on code quality.
}
\end{tcolorbox}

\textbf{Implications:} As aforementioned, the finding for RQ1 partially refutes the results of Soares et al's study, making it an interesting motivation for future studies. For instance, such studies can investigate to what extent specific combinations of refactoring types can influence refactoring effectiveness. For the practitioners' point of view, the findings of this work highlight that, despite the greater effort involved, implementing complex refactorings is highly worthwhile. This effort offers significant potential for frequently impactful improvements. Thus, the RQ1's finding also calls for tooling support and training on the application of effective, albeit complex, refactoring. 

\subsection{RQ2: SARs vs. Complexity and Effectiveness}

\begin{figure}[ht!]
    \centering
    \includegraphics[width=.9\linewidth]{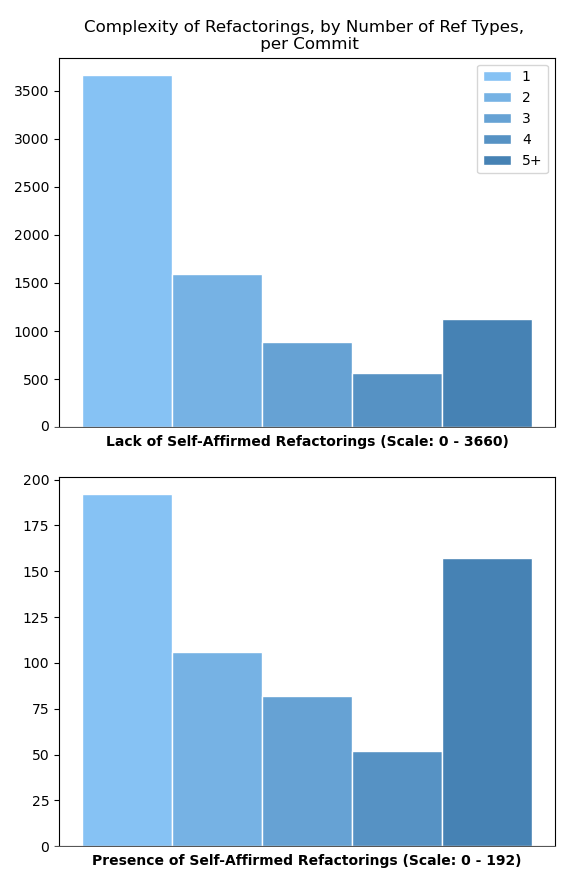}
    \caption{The frequency of self-affirmed and non-self-affirmed refactorings
composed of 1, 2, 3, 4, or 5+ refactorings.}
    \label{fig:sar_no_sar_vs_complexity}
\end{figure}

The second RQ aims to understand how the explicit intent or concern about refactoring relates to its complexity and effectiveness. Then, given our focus on explicitness, we specifically looked for the presence of SARs.
Figure~\ref{fig:sar_no_sar_vs_complexity} illustrates the frequency of refactorings categorized by complexity levels, distinguishing between those not self-affirmed (non-SAR) and those self-affirmed (SAR).

By comparing the non-SAR and SAR, we found the proportion of single refactorings to be 46.81\% and 32.6\% respectively. In contrast, we found that complex refactorings involving five or more transformation types occur at rates of 14.35\% for non-SAR and 26.66\% for SAR. This suggests that self-affirmed refactorings (SAR) are more likely to result in complex refactorings.


This finding confirms the ones from the original and other existing studies, which conclude that explicit refactorings are more frequently complex than their non-explicit counterparts~\cite{47, 8870177}. 
Using the Wilcoxon Rank-Sum Test, we found a statistically significant p-value of 0.008 (p < 0.05). However, complex refactorings (5+) were more frequent in the non-SAR dataset, likely due to using the updated RefactoringMiner 2.0.
Therefore, based on our results, we can say that when developers are concerned with refactoring to the point of expressing and documenting their concern or intent, explicitly represented through a self-affirmation of their refactoring~\cite{2}, they tend to perform more complex refactorings, spanning various refactoring types.

\begin{figure}[!t]
    \centering
    \includegraphics[width=.9\linewidth]{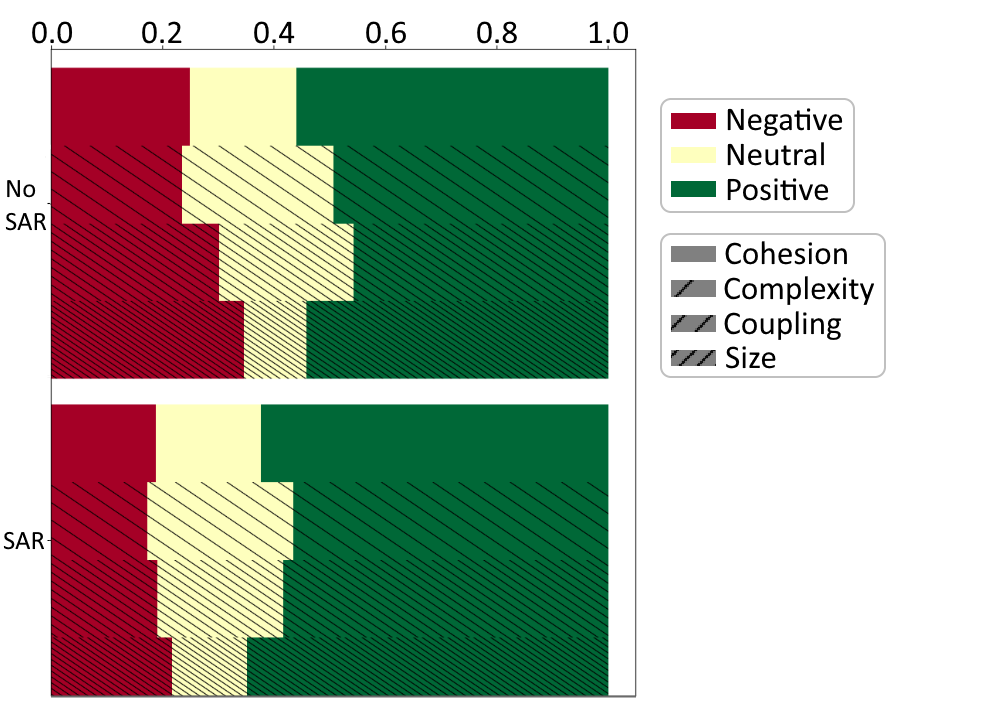}
    \caption{The negative, neutral and positive effect of self-affirmed and non self-affirmed refactorings. Each 0.1 on the horizontal scale represents 10\% in
change frequency of the corresponding effectiveness.}
\label{fig:sar_no_sar_vs_effectiveness_and_internal_quality_attributes}
\end{figure}

To answer RQ2, we analyzed the relationship between the presence of SARs and refactoring effectiveness as presented in Figure~\ref{fig:sar_no_sar_vs_effectiveness_and_internal_quality_attributes}. 
Comparing non-SAR to SAR, the average percentage of neutral changes remained approximately the same (from 20.4\% to 20.3\%) while the negative effects visibly reduced from 28.2\% to 19.1\%. This is evident from the slight increase in positive effects of SAR compared to non-SAR, as shown in Figure~\ref{fig:sar_no_sar_vs_effectiveness_and_internal_quality_attributes}. 

These results differ from the original study in that they described SARs as having usually more detrimental changes, increasing the frequency of negative effects. On the other hand, our study shows SARs as having a definite increase in positive changes, mainly reducing the frequency of negative effects. By performing the Wilcoxon Rank-Sum Test to assess the statistical significance of the results, we recorded p-values of 0.1. This indicates a lack of statistical significance. This finding contradicts the conclusions of the original work, which claimed that the results were statistically significant.

Also, Soares et al. in the original study~\cite{47}, described SARs as having a much more frequently negative effect on the code in comparison to their non-SAR counterparts. However, our study directly contradicts this finding since the frequency of negative effects was the only aspect that was visibly reduced between the non-SAR and SAR datasets.
That is, our results indicate that when developers show explicit refactoring concern, they tend to make changes with fewer negative effects, as the likelihood of positive outcomes is higher.
This aligns with \citet{8870177}, who found significant improvements in cohesion, coupling, complexity, and size metrics when developers aimed to enhance them based on refactoring commit messages.

\begin{tcolorbox}
\textit{
\textbf{RQ2 Answer:}
Our results confirm that of the original study that developers perform more complex refactorings when they make their concerns explicit along the change process. Thus, this might mean that developers write about their refactoring processes in commit messages more often when the refactorings reach higher levels of complexity. However, our results also contradict the previous work’s findings regarding refactoring explicitness's effects on refactoring effectiveness, as we found that refactorings in which developers explicitly talk about the refactoring process have usually less negative effects on the code.
}
\end{tcolorbox}

\textbf{Implications: }  The finding in RQ2 means that correlating developer
concerns and refactoring usage can be an interesting research direction for
future studies. Our findings also send at least two messages to practitioners: (i) maintain good documentation about refactoring activities to allow collaborators to uncover
and fix errors in the process before the refactoring's proponent commits, and (ii) improve the documentation of the intended refactoring may also itself directly induce the (better) quality of the refactored code.

\subsection{RQ3: NFR vs. Complexity and Effectiveness}

\begin{figure}[!t]
    \centering
    \includegraphics[width=.9\linewidth]{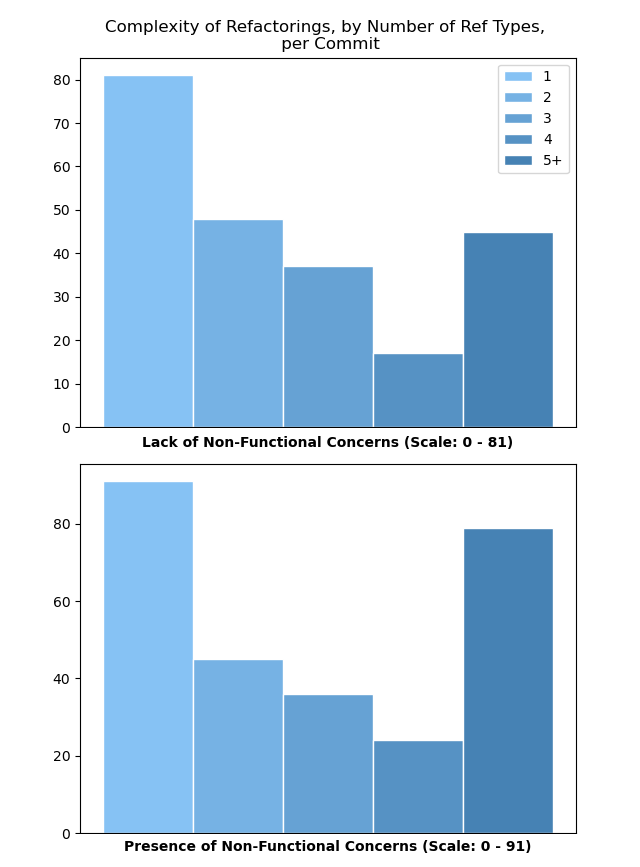}
    \caption{The frequency of refactorings composed of 1, 2, 3, 4, or 5 or more refactorings grouped by the presence of NFRs.}
    \label{fig:nfr_vs_complexity}
\end{figure}

This study aims to explore how four key NFRs (i.e., maintainability, robustness, performance, and security) influence the complexity and effectiveness of refactorings. Due to the absence of automated tools for classifying these NFRs during refactorings, we conducted a manual validation of 817 commits. Of these, 449 were identified as addressing NFRs, while the remaining 368 did not mention any such concerns. Figure~\ref{fig:nfr_vs_complexity} shows the frequency of commits for each category of refactoring complexity, given the presence or absence of  NFRs in the commits.
By comparing Category 5+ in both figures, we can see that complex refactorings are more common when developers explicitly express non-functional concerns while performing refactoring. 
By applying the Wilcoxon Rank-Sum Test between the NFR and non-NFR distributions, the resulting p-value equal to 0.91, meaning these results are far from statistical significance.

\begin{figure}[!t]
    \centering
    \includegraphics[width=.9\linewidth]{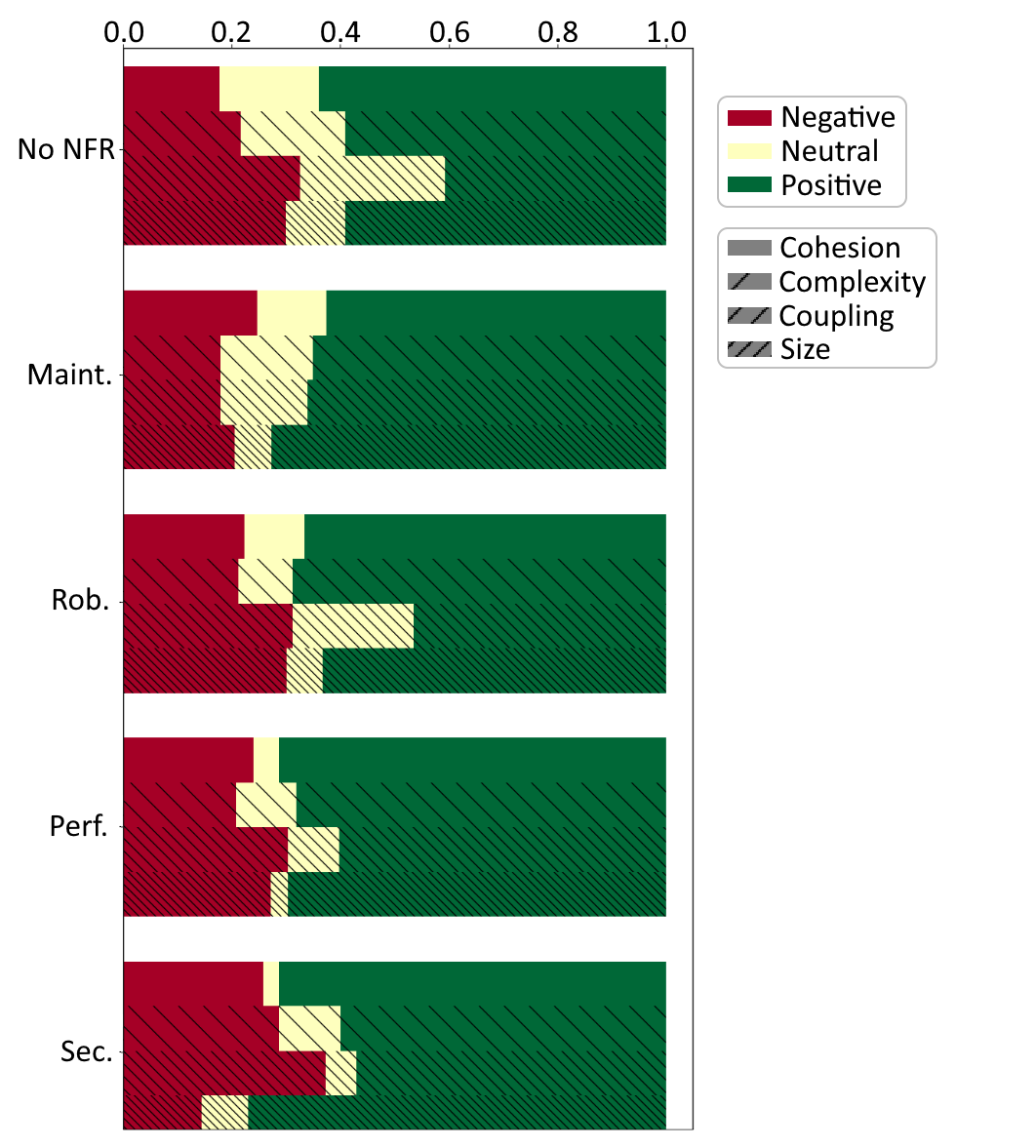}
    \caption{The negative, neutral and positive effects of refactorings when
coupled with changes in NFRs (considering only the validated data set).
Each 0.1 on the horizontal scale represents 10\% in change frequency of the corresponding effectiveness.}
    \label{fig:nfr_vs_quality_attributes_effectiveness}
\end{figure}

Also, we analyzed the potential relationship between each of the four NFRs and refactoring effectiveness, as shown in Figure~\ref{fig:nfr_vs_quality_attributes_effectiveness}.
Compared to the original work, we observed a significant increase in the positive effects of NFRs with less apparent increase in negative effects. That is, the positive effects were predominantly high relative to the negative effects. This correlates with the findings for RQ2, in which developer concerns reduce the neutral effects of refactorings, but also increase the possibility for positive effects.

From Figure \ref{fig:nfr_vs_quality_attributes_effectiveness}, we can see that refactoring in which developers were concerned with NFRs tended to have a slight increase in the frequency of positive effects when compared to the non-NFR set. Of the NFR group, refactorings with a focus on maintainability, performance and security tended to have more positive effects ($\approx$66\% of refactorings being positive on average), while robustness had a lower average (61\%), though it was still more than those without NFRs (56\%). Individually analyzing each internal quality attribute, maintainability and performance had almost the same rate of positive effects or balanced positive effects (ranging from 60 to 70\% of the refactorings being positive) across in the internal quality attributes. This observation confirms a study by \citet{iftikhar2024tertiary}, which concludes that maintainability has a moderate correlation with cohesion complexity, coupling, and size. For
robustness and security, we observed that they had more positive effects on specific internal quality attributes (ranging from 45\% to 80\% of the refactorings being positive). For instance, refactorings with security concerns had improved cohesion and code size more than the other attributes. This is in line with a study by \citet{9990968}, which observed that high cohesion, low coupling and low complexity indicate more secure software. Also, robustness led to significant improvement in cohesion, complexity and size than the other quality attributes. 

The results discussed above partially contradict the findings of the original work, which stated that maintainability and robustness had balanced positive effects across all the internal quality attributes, while performance and security were skewed toward specific internal quality attributes. In our results, while maintainability kept its balanced results, refactorings applied when developers were concerned with robustness tend to lack improvements to coupling since the combined negative and neutral effects are approximately equivalent to the positive effect. 
On the other hand, security kept its skewed results, though this time with most refactorings causing improvements to size, while refactorings applied when developers were concerned with performance had balanced results, similar to those of maintainability. Similarly to the original work, we were also unable to find complete statistical significance for the results of this finding. However, in this case, the correlations between coupling and the NFR types, with the usage of the Kruskall-Wallis test, reached a relatively low p-value of 0.07. Meanwhile, for cohesion, complexity, and size, the test reached p-values of 0.24, 0.16 and 0.33, respectively, meaning there are no statistical differences.

\begin{tcolorbox}
\textit{\textbf{RQ3 Answer :}
When developers show explicit NFR-related concerns while refactoring, they tend to perform complex refactorings at a high frequency. Also, refactorings applied when developers are concerned with NFRs have more positive effects than those in which they do not show such explicit concern.
 While we have observed that both NFRs and SARs lead to more positive effects in refactorings, SARs had less overall negative effects than NFRs. For NFRs, maintainability led to generally high positive effects across the internal quality attributes, while the other NFRs led to positive effects on specific quality attributes with less neutral effects and an increase in the frequency of negative effects.
}
\end{tcolorbox}

\textbf{Implications: } The finding in RQ3 gives the indication that mixing different NFRs into a single change may be risky as it keeps a non-negligible chance of the refactorings having an overall negative effect. Hence, developers should perform NFR-related refactoring in isolation based on each requirement and employ rigorous testing while monitoring post-refactoring effects.

\section{Threats to Validity} \label{sec:threats}
Although we performed a variety of actions to attempt to mitigate
potential threats to the validity of this work to the best of our ability, some
threats still may remain, described as follows:

\begin{itemize}
    \item \textbf{Generalizability:} While we selected a total group of eight projects, from a variety of different fields and from different developers, the results we found might still not be able to be generalized for other contexts, especially those related to closed-source projects.

    \item \textbf{Accuracy of Refactoring Detection:} RefactoringMiner, while having a high reported accuracy~\cite{53}, did not have its accuracy evaluated for our specific data set. However, other studies have used RefactoringMiner for similar sets of projects~\cite{9},  while still having a good level of accuracy in refactoring detection.

    \item \textbf{Accuracy of Self-Affirmed Refactoring Classification:} The keyword-based classifier we used for SAR classification in this work may not be generalizable to other projects, due to the keyword set potentially needing to change. However, to mitigate this, we have performed an evaluation of the classifier's accuracy with regard to a manually validated sample of the data set, leading to a decently high level of accuracy, and still surpassing the accuracy of another, state-of-the-art classifier in this specific dataset.
\end{itemize}


\section{Conclusion} \label{sec:conclusion}
In this study, we have replicated the study by Soares et al.~\cite{47} with an extended dataset composed of eight different open-source software projects. Our focus is to validate the findings of the original study and further understand the relation between complexity (number of refactoring types applied),  effectiveness (level of improvement in internal quality attributes), explicitness (presence of SAR), and NFRs related to refactoring. 

The results from our analysis provide valuable insights. Firstly, our results confirmed that refactoring complexity is linked to its effectiveness and that as the complexity increases, the effectiveness of the refactoring increases as well. 
Secondly, we found that developers tend to apply more SARs, and that SAR is generally more complex than their non-SAR counterparts. Similar to that of refactoring complexity, we found SARs to be more effective in improving internal quality attributes than non-SARs. However, this contradicts the original study in that they observed SARs to have detrimental effects on quality attributes.
Lastly, we found that when developers express explicit concerns with NFRs while refactoring, it results in complex refactorings that have positive effects on internal quality attributes. Considering each of the NFRs we considered in this study, we found that maintainability and performance had a balanced positive effect. In contrast, security and robustness had a skewed positive effect toward specific internal quality attributes. We found this to partially contradict the findings from the original work, which mentioned Maintainability and Robustness had balanced positive effects.

Our findings offer contributions for both developers and researchers.
For developers, our results highlight the importance of maintaining thorough documentation on refactoring activities and their underlying intents. Additionally, they emphasize the need for caution when applying refactorings that may impact non-functional requirements (NFRs), as certain NFRs can degrade specific code quality attributes.
Future research should explore the trade-offs between improving internal quality attributes and satisfying NFRs during refactoring, as well as the effects of different refactoring combinations to better understand their effectiveness and mitigate negative impacts.


\section*{Data Availability}\label{sec:data}
Artifacts of the evaluation are available online~\cite{soares_2025_14780166}.


\balance
\bibliographystyle{ACM-Reference-Format}
\bibliography{references/bibtex}


\end{document}